# Semantic Modeling of Analytic-based Relationships with Direct Qualification

Norman Ahmed
Purdue University and AFRL/RI
Department of Computer Science
West Lafayette, IN 47096
ahmed24@purdue.edu

Jason Bryant, Gregory Hasseler, and Matthew Paulini
Air Force Research Laboratory/RI
Rome, NY 13441
{jason.bryant.8, gregory.hasseler.2,
matthew.paulini.1}@us.af.mil

*Abstract*— *Successfully modeling state and analytics-based semantic relationships of documents enhances representation, importance, relevancy, provenience, and priority of the document. These attributes are the core elements that form the machine-based knowledge representation for documents. However, modeling document relationships that can change over time can be inelegant, limited, complex or overly burdensome for semantic technologies. In this paper, we present Direct Qualification (DQ), an approach for modeling any semantically referenced document, concept, or named graph with results from associated applied analytics. The proposed approach supplements the traditional subject-object relationships by providing a third leg to the relationship; the qualification of how and why the relationship exists. To illustrate, we show a prototype of an event-based system with a realistic use case for applying DQ to relevancy analytics of PageRank and Hyperlink-Induced Topic Search (HITS).*

*Keywords-component; Semantics, Semantic Technologies; Knowledge Representation; Analytics; Direct Qualification.*

## I. INTRODUCTION

Two common semantic modeling perspectives are data-centric and knowledge-centric. A data-centric view of information can be simplified to the management of archived documents or publication events. A knowledge-centric view of information models the content of data sources as an intersecting map of facts and assertions. By embracing an aggregate perspective that combines relationships for unstructured knowledge representation with structured, document-centric relationships, the process of determining, modeling, and expressing relevance with semantic technologies can be performed. However, combining multiple semantic modeling perspectives increases the complexity of the resulting model.

Generally, semantic relationships consist of a subject/object pairing and a directional predicate. For quad-based relationships the context is added by organizing triples into a named-graph set. The weaknesses of this design include the inherent complexity of domain ontology modeling, the rigidity of modeling rules, and the granularity of semantic entities and the predicates. For example, when these predicates are not suitable, a new one is constructed. Since two predicates with similar meaning can only be equivalent or in-equivalent, thereby, lending to ineffective relation linkages. To the best of our knowledge, there is no mechanism for expressing the manner in which a relationship linkage is qualified. This is where Direct Qualification (DQ) provides a third relational leg to provide provenance to the previously unqualified, unjustified, and trust-agnostic relationship.

In practice, semantic state representation challenges are a side effect of integrating semantic technologies with distinct areas of focus; one being knowledge representation (RDF), the other being ontology instantiation (OWL). OWL can be seen as the semantic web equivalent of schemas to the standardized document object model. Semantic relationships involving fluctuating values cannot be modeled with sufficient granularity using existing semantic standards or technologies.

Attempts to mitigate this shortcoming have focused on semantic knowledge representation scoping, link analysis, broadening query expressions, manipulating query results, and natural language processing. Unfortunately, such attempts typically encounter issues of *reification*, the consequence of attempting to simplify all relationships into Subject-Predicate-Object sets within semantic technologies. Other issues include; many interrelated values, lack of standards-based provenance, or ontological versioning conflicts. However, these hindrances can be avoided by using finer grained relationships of provenance. Provenance can, in a basic form overcome through the adoption of named graphs instead of pure triples. Named graphs enable a simple but foundational level of document sourcing and traceability. Resolution of conflicting versions of ontological entities and predicates can be overcome through adoption of standard "Source/Domain/Version" structures for URIs (e.g. http://mysourcenamespace.com/stateontology/V2.0/). While these workarounds are straightforward, they increase complexity while decreasing maintainability and performance without the benefit of supporting advanced semantic features.

The purpose of DQ is to provide context to the meaning of established, semantic relationships, such as relationships to events representing results of analytical algorithms over a graph [11].

In this work, we show that Direct Qualification (DQ), an application of the Prov-O ontology, can be paired with Relevancy Ontology to model state, including that of analytical relationship context, in a way that avoids *reification*, an intrinsic complexity inherent of the semantic standards. This ability allows us to apply generic, graph-based analytics to semantic knowledge and model the results. We demonstrate the effectivenes of our approach by applying the *Betweenness*, *PageRank*, and *HITS* algorithms to a semantic-based knowledge graph derived from a realistic scenario. Thus, our main contributions can be summarized as follows:

- Introduced mechanisms for any semantically referenced document, concept, or named graph to be associated with the results of applied analytics.
- Enabled the `ORDER BY` command of SPARQL to construct a document result set graph sorted by any expressed, normalized analytic score.
- Created the ability to model analytic provenance and analytic results, allowing either to be reasoned upon.
- Enabled determinations of efficiency for different analytics and the potential to combine analytic-based queries with semantic queries.

We have organized the paper as follows: We first give a brief background of the subject and the motivation behind our work in Section II. We then introduce DQ and discuss the details of our approach in section III, and its limitations in Section IV, followed by the design and implementation in Section V. We consider the experiments in Section VI, followed by the related work and conclusion in Sections VII and VIII, respectively.

## II. BACKGROUND AND MOTIVATION

Semantic technologies are well suited for modeling explicit and fully qualified relationships but ill suited for state full relationship entities that can change overtime. The ability to model the results of analytic algorithms in a way that preserves the integrity of the fundamental knowledge in the semantic graph is critical. For example, SPARQL queries could be paired with graph-based analytics that have been executed on a particular semantic-based knowledge graph and document-based analytics available from other query languages and platforms.

These stateful data attributes need to be represented either ontologically as an entity instance in which has the potential to explode the quantity of relationships, or as an unstructured relational fact defined without an OWL counterpart which exists in a vacuum of provenance. Applying these observations to an realistic enterprise use case demonstrates the complexities of mapping abstract solutions to concrete semantic problems. Resource examples include entities such as a document, a book or website instance, an analytic execution result, a knowledge extraction, or a formalized representation of an e-mail message. Semantic facts expressed within these entities could be a name, current geo-location, condition status, social POCs, or target details. These examples result in two alternate approaches to semantic state management; present and historical relation updates.

*A.* **RelationshipUpdating:** A single semantic instance of a resource is updated with the 'present' relationship value whenever it's adjusted. Implementing this scheme is simplistic and results in a constantly up-to-date semantic model. However, for most realistic domain use cases,this is seriously flawed. For example, it excludes all past state changes for a resource's dynamic relationships. By replacing old state values with new state values, it removes all capacity for tracking non-static relationship values; past state can never be queried and, consequently, never learned from, such as with trend analysis.

*B.* **RelationshipVersioning**: Maintaining a historical record by creating a new instance, or concretization, of the resource and its present relationship states. Over time, particularly if an event has relationships that change quite often, numerous of versioned instances of the same event could be created, making queries overly complex and resulting in a high degree of overhead due to duplication for relationships may or may not be static.

Provenance for semantically represented documents provides support such as: attribution of authorship, change tracking, sourcing, and transformation of; entity, activity, or agent involved with the document. However, semantic provenance for documents supports a slightly different subset of features than it does for those which are non-document-centric. The Non-document-centric relationships lack the context of sourcing, unless explicitly and independently associated via provenance relationships. Consequently, semantically-expressed relationships lack traceability to sourcing and thereby lose traceability to authorship, time of creation, and other common provenance attributes.

Furthermore, semantic provenance for documents is handled less explicitly, by using the unique document URI as the named graph, or quads, for all relationships extracted from the document. This traceability is important because without introspection into referential sourcing there is a limited measure of trust in data content. For non-deterministic relationships, the implications of using semantics grow even more complex. Many probabilistic analytics improve reliability and trust when trends of an increasing or decreasing score are tracked over time.

Additionally, semantic-based graphs traditionally only support logic-based reasoning, whereas non-semantic graphs are commonly evaluated using traversal queries, popularity, similarity, clustering, or other externally executed analytics. Analytical graphs have never been sufficiently modeled ontologically, in order to enable pairing with semantic inferencing engines. DQ is the means by which semantic modeling can be applied to express these probabilistic or analytics-based relationships, bridging the gap between semantic inferencing logic and graph-based analytics.

A key focus of this work is to enable new features for semantics-based modeling of relevancy scoring and analytics. In addition, provide a formal method to represent analytic provenance, enable the persistence of historic scoring records, and expand the range of inferencing capabilities of the semantic repository and executed queries.

## III. DIRRECT QUALIFICATION

Direct Qualification (DQ) is an application of the W3C Provenance Ontology (Prov-O) [12] that can be paired with Relevancy Ontology to model state and analytical relationships of documents. Such pairing scheme enables modeling the results of generic graph-based analytics to enhance semantic knowledge.

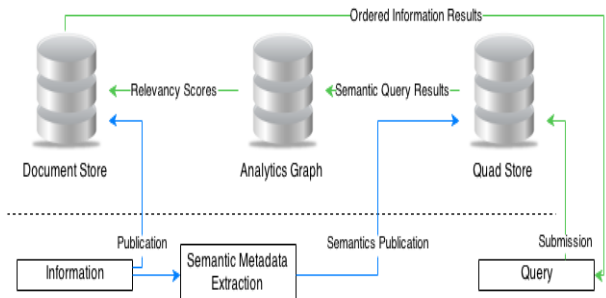

*Figure 1.High-level DQ Architecture*

As depicted in Figure 1, the *document store* (left cylinder) traditionally supports content-based queries, the *analytics graph* (middle cylinder) supports graph-based analytics and queries, and the *quad store* (right cylinder) supports semantic inferencing and rule logic. Our DQ approach enables content-based and trend-based analytical queries within the *quad store* by representing the ontologies and models necessary for bridging the graphs. In other words, the semantic graphs and non-semantic graphs become interoperable, enabling any combination of rule logic, inferencing, and graph analytics.

Therefore, DQ not only enables query capabilities for both the analytical results and the captured analytic qualification relationships for provenance, it also provides for semantic queries with information that was previously non-existent, non-queryable, or incorrectly deemed irrelevant. However, key challenges include how to semantically represent the relationships that change overtime, or s*tateful semantics* (discussed in section A); and how their *relevancy relationships* (discussed in section B) can be represented while preserving the W3C standards that these models are based on (discussed in section C).

### A. Stateful Semantics

Semantic technologies have generally been used to represent the categorization properties of a domain of knowledge over which some form of rule logic and learning can be executed. Most semantically represented bodies of knowledge involve the declaration of data properties that tend to be evaluated on their existence rather than on their propensity to change. Thus, the notation of state in semantic technologies is largely ignored.

In the most standard use case, that of content-based document modeling (World Wide Web paradigm), documents change are infrequently and, when they do, it can be considered a new version of the old document. The focus is on inference and reasoning for determining truths or, at a minimum, to monitor events. For some documents like reports or books, iteratively improved content is acceptable. For others, the only association between old content and new content may be the source URI such as Newspapers, Facebook news wall, or a search engine front end, or graphics [3] etc.

Semantic resources, similar to documents, have an intrinsic identity with possessive traits. The distinct difference, however, is that the semantic resources are much more likely to have attributes that change state, which should never reflectively alter the identity of the possessive asset. For example, consider a newspaper as an iterative publication with the same title at the top each day but fundamentally a new instance with static content. In the case of a web-based newspaper, there is no temporally canonized version, rather a fluctuating set of aggregated news stories throughout the day which, incidentally, do not alter the identity or name of the newspaper itself. Thereby, it emphasizes an iterative document-centric versioning paradigm.

On a different prospective, consider a scenario for a semantic representation of a person, with its various possible relationships (e.g. height, weight, medical status, and name). If a person's weight and medical status is altered daily, relationship duplication for thousands of people would expand dramatically in a short period of time. Similarly, a resource with states changing almost constantly, such as a network router's traffic, could result in tens of thousands of status changes hourly. While this view maintains traceability, enabling semantic queries of both present and past relationships, it overburdens storage and inferencing resources, as well as overcomplicates the query process by treating state changes as duplication triggers. All queries for an event would be required to disentangle past and current instances. The key relationship used to manage stateful data is the `specializationOf` predicate of Prov-O which is intended to apply state-based relationships to any Entity or an Agent.

### B. Relevancy Representation

Representing the relevancy of documents measured via analytics, that are scored either independently or in concert with other documents, presents multiple obstacles. First, the main obstacle is within the provenance modeling itself. Modeling document relevancy should result in a cohesive solution that utilizes a clean, simple, and straightforward query expression. However, many provenance-based qualification relationships apply to use cases where entities, activities, or agents have been qualified either singularly or in pairs, but not as a combination. This greatly increases the complexity of the query expression. Second, the complexity of reification, which is the consequence of attempting to simplify all relationships into Subject-Predicate-Object sets within semantic technologies, makes it difficult to express the results of an analytically-qualified relationship between entities with cardinality greater than one.

Analytics that can process a semantic graph and enable relevancy determinations must be applied selectively through best-fit evaluations. For example, some analytic frameworks, such as Google's MapReduce or Natural Language Processing, operate over raw documents upon either their publication or their returned DB query result. These frameworks are tailored for document or key/value pair stores, such as Hadoop or Cassandra. Semantic inferencing can create a form of analytics by applying an ontology relevant to the domain and setting up type, sameness, and equivalence rules.

For example, if an extraction from two documents results in the determination that there are matching identifiers for an object or person, then the respective attribute can be inferred to have "*sameness;*" instance equivalence. This is a common feature of establishing property chains within OWL-2. Semantic reasoners support rule-logic that can determine if the value of a relationship has reached a particular threshold but do not support analytical features for graph traversal, modeling, or state change events. This is where our DQ approach fills the gaps.

Determining relevancy provides relationship models for information quality, prioritization, and results of stochastic algorithms for information systems. Some forms of relevancy can be compared to enhancing a search engine like Google [7]. Documents found by keyword searches have long been ordered by estimating relevance through application of "popularity" analytic. This has generally remained unobtainable for semantic queries because the use cases have either not been document-based or were unable to be processed by analytical algorithms that don't use rule-logic within semantic data sets.

The internet is an HTML representation of multiple knowledge domains overlayed upon a set of segmented documents with unique URLs, while a document-oriented semantic dataset is an RDF/OWL representation of multiple knowledge domains overlayed upon a set of segmented named graphs with unique URIs. Both datasets contain interlinking between documents; the only difference being that the links on the internet are not explicitly tied to a particular ontology. Many web-oriented popularity, similarity, and clustering analytics appear to be well suited for semantic datasets. Our proposed DQ framework adopted an initial set of analytics for use, including PageRank, HITS, and Betweenness.

## C. Semantic Standards and Direct Qualifications

In this section we discuss the challenges involved in applying DQ while complying with the existing semantic standards, W3C Provenance Ontology (Prov-O), and Relevancy Relationships. The purpose of DQ is to provide context to the meaning of established, semantic relationships that represent results of analytical algorithms over a graph. Key concepts of the standards involved in this context include: *qualification, provenance, specialization, relevancy scoring, occurrent relationships and continuant state events, and finally, monotonicity.* We briefly outline the idea underlying each of the concepts and discuss how DQ extension schemes are applied.

### 1) QUALIFICATION

The qualification of a relationship is an expression of how two entities or, in our case, documents are associated. For example, the Friend-of-a-Friend (foaf) ontology supports properties that can declare whether two people know each other but does not support the qualification of how, why, and to what degree they know each other. They also do not capture probabilistic relationships, such as the result of an analytic determining a strong likelihood that one person knows another.

DQ provides the discoverability and traceability of the analytical qualification that result in the probabilistic relationship. Possible relationship qualifications that could be applied to the `knows` property of the foaf ontology are found in Prov-O. With DQ, the provided set of qualifications is general enough to provide generic use case solutions but specific enough to extend domain-based qualifiers, if necessary.

### 2) PROVENANCE

Provenance is the mechanism that enables data or information to be monitored or traced back to its origin or history, such as how the data is created, transformed, or authored. Such mechanisms offer a powerful means by which to enact trust, quality metrics, and security measures.

Prov-O, currently a W3C recommendation [12], seeks to provide a set of general provenance concepts and properties for interconnecting entities, activities, and agents. As an example, some supported properties are `wasAttributedTo`, `wasAssociatedWith`, `used`, `wasGeneratedBy`, `wasInformedBy`, `actedOnBehalfOf`, and `wasDerivedFrom`.

Our work extends the provenance properties to create probabilistic qualifications within a relevancy ontology, expanding the range of those within Prov-O to a broader set of domain relationships. The Prov-O concepts of `entity`, `activity`, and `agent` are extended to relevancy counterparts such as *idempotent*, *Stochastic*, or *Boolean analytics*. These enhancements are intended to support modeling of analytic qualifications and the resultant attributes.

### 3) SPECIALIZATION

Provenance supports relationship state changes with varying degrees of complexity, via specialization. Specialization qualifications within Prov-O reduce relationship duplication while creating the capabilities for more advanced analytics. Semantic technologies serve multiple knowledge representation use cases, one being the expression of static relationships (*Occurrent*), the other being the expression of stateful (*Continuant*) relationships, discussed in sections 5. Either one of these relationship types may be expressed independently as an RDF triple, or message, or dependently as a an RDF quad, or document, , using the source message / document as a possessive named graph.

RDF and OWL are ideally suited to stateless and independent facts [13], as long as trust determinations, document traceability, and sourcing are non-critical features. However, in most real world use cases, these features become critical for determining truth and relevancy. For example, if semantic relationships are extracted from multiple sources disagree on a date, identity, or other relevant facts, then these facts are expressed solely as triples, a determination of truth must be made as to which relationship is correct. However, determining truth among conflicted relationships is nearly impossible without first determining trust levels of the extraction sources. Once a source becomes more trusted than another, a form of provenance is pulled back into the process, whether implicitly or explicitly.

In addition, relationship types can have an impact on selecting the ideal modeling solution. Some attributes of an asset may be occurrent (e.g. name, identity, asset type), while others are continuant (e.g. fuel level, latitude, longitude, role). Semantics intuitively treat all relationships as *occurrent*, with minor allowances for limiting their cardinality if customized to a domain. OWL, SPARQL, and most other ontologies do not have a built in mechanism to support the distinction between *occurrent* and *continuant* relationships. In order to retrieve changes of state for a data or object attribute, that attribute must be explicitly defined within the ontology or a customized additional layer of abstraction.

*4) RELEVANCY SCORING*

Information can be analytically scored independently, co-dependently, or in combination. Additionally, any modeling solution needs to support different algorithmic processes determining whether the analytic result is numerically-based or Boolean, as well as whether the result can be normalized as *idempotent*, or is measured against a probabilistic threshold. DQ supports these complex and diverse analytical models, since Prov-O and relevancy ontologies can be extended to niche analytic domains. DQ can also support future collaborative analytic ontologies.

For example, let's consider stochastic analytics that result in a non-deterministic value for the state of a relationship. Calculating the Vector Space Modeling (VSM) [4] similarity of two documents may result in a particular keyword being correlated by a value, let's say 5.6. Is this important or relevant? It lacks meaning unless normalized and measured against a threshold acting as a heuristic determination. We will show, in the experiments section, a subset of the analytics executed via the semantic graph, including SPARQL queries and relevancy analytics for "*Betweenness*", "*PageRank*", and "*HITS*".

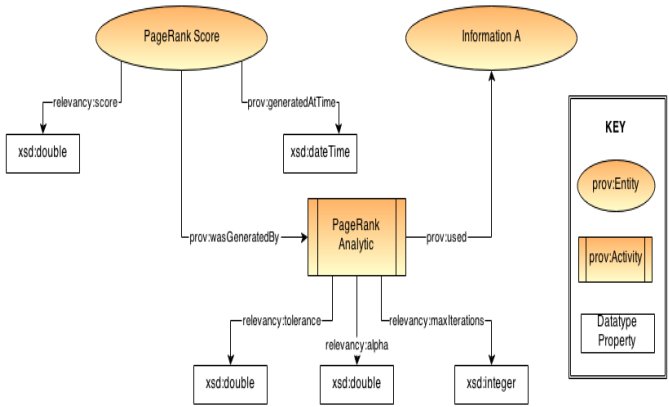

Figure 2: PageRank Analytics with DQ

*5) OCCURRENT RELATIONSHIPS*

Occurrent relationships present the simplest opportunity to apply DQ. They are relatively simple to model, will never require adjustment, and have a common form of expression, regardless of the type of analytic being applied. Figure 2 illustrates the application of the PageRank analytic to a document. The PageRank analytic applied to semantic node graphs produces dynamically-adjusting, normalized scores because the publication of subsequent documents increases the amount of interconnections for the document. This should, theoretically, predict an increased likelihood of popularity.

Such analytic differs from others, such as the VSM estimation for a single document as depicted in Figure 3 below. Even if VSM estimates are aggregated for an entire collection of documents, the score will not change based on alterations within the knowledge graph. Regardless of subsequent published relationships, VSM will result in the same normalized values of keyword frequency. These aggregate document analytics, however, leave the monotonic realm of entity-to-entity DQ models and enter the realm of entity-to-entities DQ models. The two complex use cases for applying DQ is continuant relationships that derive from applied analytics and another is for qualification relationships that are non-monotonic.

*6) MONOTONICITY*

Monotonicity has different meanings depending on whether it is applied in mathematical functions, logic, or semantics. In this work, we use the term to describe the unidirectional, single-object capabilities that limit semantic relationships.

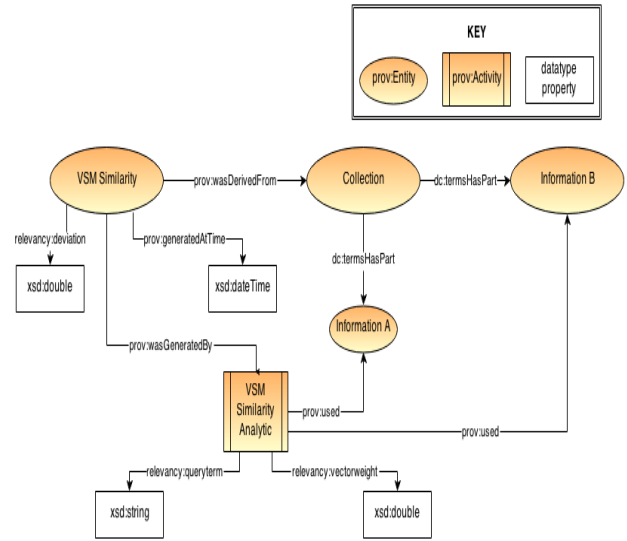

Figure 3: VSM Analytics with DQ

Figure 3 above illustrates a non-monotonic relationship applying DQ between multiple documents. This is a complex case because DQ is being utilized as an alternative to reification in qualifying an analytic to multiple sources or causes. Applying DQ in this form to analytical values, results in the capability to semantically express the provenance, inputs, qualifications, and results of analytics; together these can validate the quality and relevance of the information.

## IV. DQ LIMITATIONS

DQ may not be necessary for all cases where analytic provenance is desired. Applicability can be determined by applying the decision matrix in Figure4.

|  | State Change Event | |
|---|---|---|
| Relationship Type | Continuant | Occurrent |
| Relationship | Specialization and Direct Qualification | Direct Qualification |
| Attribute | Specialization | Basic Semantic Inferencing and Reasoning |

DQ is applicable for all analytics-based value outputs, although it is important to note that DQ is implemented differently depending upon the nature of the event or data/system types. As depicted in the matrix decision table above, in the case where semantic data or object relationships are expressing an event that is occurrent and non-probabilistic in nature, DQ modeling will not yield any new insights over traditional semantic inference or reasoning. In the case where an entity-entity relationship is continuant, or state-based, but non-probabilistic in nature, the specialization provenance feature enables traceability for state management.

## V. SYSTEM DESIGN AND IMPLEMENTATION

The high-level system architecture illustrated in Figure 1 (section III) depicted a set of three data stores (cylinders) document store, a non-semantic analytics graph, and a semantic quad-store. Unlike a file-based system, the data flows are orchestrated to support document publication and streaming deliveries. Ontologies supported by our framework include the common solutions for time, geospatial (GeoSPARQL), common elements (U-Core SL), and mission planning (Cornerstone), with a custom ontology for publish and subscribe middleware. Additionally, before the data can be extracted and characterized, there is a pre-processing stage consisting of format determination and XML type determination (e.g. schema, domain specific XML message type, XML-MTF types), if applicable.

### A. SYSTEM COMPONENTS

In this section we describe the technologies leveraged to build the proposed scheme. DQ required the construction of supporting infrastructure, including a quad-store, a raw document database, and a graph-based analytics framework. We leveraged widely adopted, open-source technologies:

- Jung, an open-source, graph-based analytics framework, was extended to enable its built-in analytics over semantic graphs represented using the Jena framework interfaces.
- Parliament and Virtuoso used as quad-stores.
- Aperture, an open-source mime-type determination and content extraction library, was used to resolve the format of messages in order to apply the appropriate extractors.

In addition, we fully developed an extraction services in our previous work [1], for DQ. For those interested the details, a more verbose discussion of the design and implementation of the framework given in our in-house technical report [2].

### B. ALGORITHMS

The primary steps for enabling Direct Qualification are as follows:
i) Support persistence of cross-referenced, raw documents and semantic quad-based relationships by using the semantic URI of the document's named graph as the unique key for the raw document retrieval.
ii) Strictly enforce the separation of the semantic models for class instances from the events affecting their state relationships.
iii) Support graph-based processing of analytics over semantic edges and vertices by bridging semantic and analytic graphs.
iv) Support event-based relevance scoring triggers, such as SPARQL queries, XPath queries, metadata extraction, semantic reasoning, or keyword searches of raw text.
v) Determine the appropriate Direct Qualification Model based upon tests for occurrence, continuance, and the monotonicity of the entities involved in the applied analytic.
vi) Express the relevancy qualifications and scoring of documents through the pairing of standardized provenance ontology with Relevancy ontology.
vii) Persist the DQ results within the quad-store.

## VI. EXPERIMENTS

In order to illustrate support of stateful relationships, we developed a lightweight, event-based information share using publish, subscribe, and query middleware model. Key participants in the system are a publisher, a subscriber, and a broker. Sample publishers and consumers were established, as well as the implementation of a quad-store (Parliament), an analytics graph system (Jung), and a raw document database (Hash Map with file references).

Simulated geo-location data for tracking cell phones that are variably repositioning was published and semantically expressed via a set of indexers. The result of semantic processing is an RDF/OWL document that relates details involving times, point-of-contact (POC), status, and latitude and longitude. Each document is viewed concurrently as an independent publication with an associated semantic named graph. Each named graph has, at a minimum, the following fields:

- Named Graph URI
- Information URI
- Publisher Identity URI
- Publisher Role
- Message Topic
- Message Type
- Message Format
- Time Published

Optional fields exist for a wide range of relationships, including:
- POC Involvement
- Resource Involvement
- GeoSPARQL Compatible Geolocations
- Keywords
- Publisher Geolocation

After publication, the quad-store has instances defined for all resources and locations involved but no continuant state changes. After pre-processing, format determination, type determination, and semantic extraction completes, the analytics and DQ are executed as part of the query process.

Once the semantic and non-semantic graphs were made interoperable via Jung and Jena with bridging code (within Java source), we constructed an experimentation harness that performed sample SPARQL queries and relevancy analytics over the test scenario. The initial set of analytics applied included "Betweenness", "PageRank", and "HITS".

Each document is viewed concurrently as an independent publication and a semantic named graph, while the query results are autonomously mapped from a semantic graph result set to a traditional Vertices/Edge graph tailored for analytics scoring. After the pre-processing completes, state-based queries are enabled for any new observations of state. An advanced set of query capabilities can be enacted over this data. This allows state traceability and discoverability of the newest particular state of an asset. These are key features that were previously not available in the semantic domain.

### A. EXPERIMENTAL SETUP

Experiments were run on Commercial-off-the-Shelf, quad-core desktop machines with 8 GB RAM and Windows 7. As the scenarios were intended to demonstrate a proof-of-concept rather than scalability or performance metrics, neither cloud-based nor high-performance server machines were required.

The analytical proof–of-concept test scenario consisted of 230 real messages published over a period of 10 minutes. The message types included geo-location messages with status updates for cell phone usage. The semantic relationships created were produced by means of the extraction framework.

### B. EXPERIMENTAL RESULTS

We first show the results of applying analytics with DQ.

| Impact Information | HITS Initial Score | HITS Final Score | Increase % | Total Increase |
|---|---|---|---|---|
| Information 7 | 0.2404077 | 0.2919673 | 21.45% | 0.0515596 |
| Information 16 | 0.24256352 | 0.29383602 | 21.14% | 0.0512725 |
| Information 17 | 0.24233185 | 0.29263605 | 20.76% | 0.0503042 |
| Information 11 | 0.24166479 | 0.29196729 | 20.81% | 0.0503025 |
| Information 20 | 0.24278861 | 0.29196731 | 20.26% | 0.0491787 |

Table 1: HITS Results

Table 1 illustrates the PageRank scoring throughout the life of the experiment. As shown, some information increased their levels of inter-relations with other entities at a higher rate than others. Just as in web documents, PageRank can show which named graphs are increasing in estimated popularity. The PageRank analytic is resampled for all messages after every 10 message publications. Additionally, the HITS table, Table 2 shows a similar set of measures, including an average increase in popularity, starting from the initial publication and ending with the final.

| Impact Information | PageRank Initial Score | PageRank Final Score | Increase % | Total Increase |
|---|---|---|---|---|
| Information 6 | 2.50171052 | 2.58682592 | 3.40% | 0.0851154 |
| Information 122 | 2.17891748 | 2.203650104 | 1.14% | 0.024732624 |
| Information 109 | 2.17586335 | 2.189030325 | 0.61% | 0.013166975 |
| Information 135 | 2.72118958 | 2.736303557 | 0.56% | 0.015113977 |
| Information 118 | 2.1794559 | 2.189030328 | 0.44% | 0.009574428 |

Table 2: PageRank Results

Our proof of concept for DQ seeks to apply relevancy analytics that are intended for the world-wide web to semantic analysis. Our research does not validate the quality of these analytics; it only seeks to demonstrate the concept for semantically modeling them appropriately. PageRank attempts to determine relevancy based upon links between webpages or semantic named graphs, utilizing weight normalization to determine the likelihood of being at the node while doing random graph traversals. HITS determine the degree to which the named graphs are mutually reinforcing and which named graphs are authorities (highly interconnected hub).

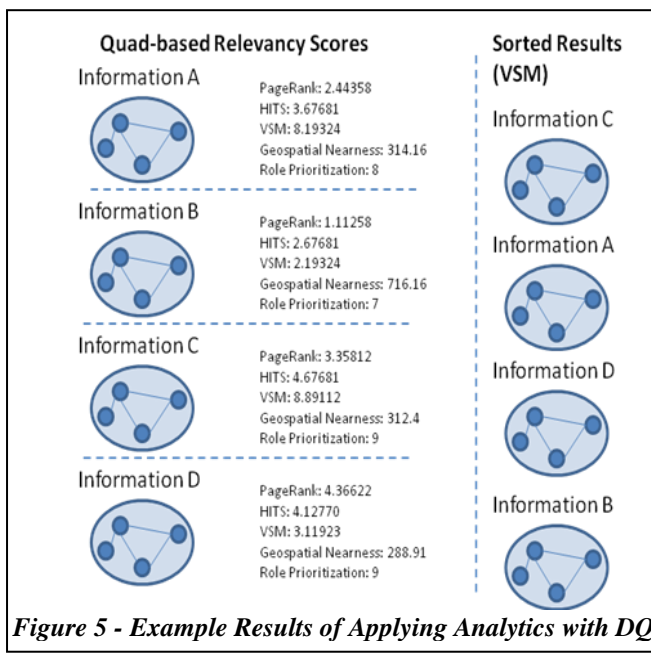

*Figure 5 - Example Results of Applying Analytics with DQ*

Figure 5 above show DQ results for analytics applied to semantic named graphs, illustrating multiple analytic values for a query result set; these are what would be published to the semantic quad-store. A key advantage of pairing provenance and relevancy vocabularies is the power it provides for overcoming the complexity of semantic reification. Reification is normally implemented when a semantically modeled instance is seeking to express either the qualification or provenance of a relationship. These two cases can be mitigated without resorting to reification, however. Adopting a quad-based perspective of semantic relationships can achieve a basic form of provenance by allowing traceability to the source named graph's unique URI.

Prov-O expands the set of provenance support and supplies some generalized predicates for qualification. This effectively solves the non-probabilistic subset of analytical use cases. Even pairing both of these approaches, there is a failure to solve the qualification of probabilistic analytics, such as the results of Vector Space Modeling, PageRank, HITS, or other Natural Language Processing analytics. This failure occurs because the only supported provenance attribute becomes the source URI. Although the source URI could be related to analytic attributes and provenance qualifiers, the model would still not support probabilistic analytic relationships. Our DQ approach limits the provenance deficiencies, while supporting the extensibility of domains and ranges for analytic relationships.

## VII. RELATED WORK

As semantic standards mature and applications expand into new domains, research regarding semantic management of stateful relationships is beginning to be explored more fully. Current research has been tangential, at best, while missing many of the niche problem areas of semantics. Approaches in this area have focused on inferencing through the use of join sequences or resolving models with conflicting states [9]. To wit, approaches involving applied analytics for state management have attempted to do so during the extraction phase of data processing [8], rather than utilizing semantic technologies or ontology models.

While there are ongoing efforts towards document analysis using analytics such as PageRank and HITS [5], and VSM [10], the focus of those efforts has been on ontology matching [4] or temporal/geospatial query enhancement [6]. Our approach differs in that it stays confined to semantic technologies with special emphasis on event-based information sharing, modeling, data mining, and retrieval, all combined. Furthermore, one of the key differentiator of some existing semantic models with DQ approach is that they adopt a constantly "present" based view that updates the instance with relationships reflecting any changes in its state. Thereby, the state changes' value can never be considered truly distinct from its identity URI.

## VIII. CONCLUSION

Traditional semantic data model approaches fall short when confronting the challenge of state-based relationships. They focus on static knowledge representation, extractions of static data properties, or enabling information management features via rule engines and inferencing. Our solution provides a new layer on top of traditional approaches, with independently defined events to reconcile integration conflicts using DQ.

Managing the state of a published information event significantly reduces the computation time of semantic queries, the load on semantic DBs, and elimination of wasteful property and instance duplication. Furthermore, it enables advanced query heuristics, makes state change instances more lightweight, and organizes state changes temporally so that the newest events are easier to discover.

In our future work, we explore semantic state traceability paired with semantic graph analytics. Reasoning over stateful trends within segmented time periods can demonstrate possible advanced uses of semantics for stochastic and Boolean-based analytics applications, thus producing support for prioritization, query result set ordering, and provenance modeling of analytics.

### ACKNOWLEDGMENTS

The authors would like to thank Tim Lebo for his valuable semantic discussions in the initial design of our prototype and the Phoenix [2] development team at AFRL for the traditional pub/sub system that provided the foundation for our prototype.